\PassOptionsToPackage{hyphens}{url}

\documentclass[sigconf]{acmart}
\copyrightyear{2024}
\acmYear{2024}
\setcopyright{rightsretained}
\acmConference[CCS '24]{Proceedings of the 2024 ACM SIGSAC Conference on
Computer and Communications Security}{October 14--18, 2024}{Salt Lake City,
UT, USA}
\acmBooktitle{Proceedings of the 2024 ACM SIGSAC Conference on Computer
and Communications Security (CCS '24), October 14--18, 2024, Salt Lake City,
UT, USA}
\acmDOI{10.1145/3658644.3691387}
\acmISBN{979-8-4007-0636-3/24/10}



\usepackage{threeparttable}
\usepackage[utf8]{inputenc}
\usepackage{graphicx}
\usepackage{subcaption}
\usepackage{fancyvrb}
\usepackage{multirow}
\usepackage{makecell}
\usepackage{array, multirow}
\usepackage{arydshln}
\newcommand{\ignore}[1]{}
\usepackage{balance}
\usepackage{xcolor}
\usepackage[ruled,vlined]{algorithm2e}
\usepackage{booktabs}
\usepackage{listings}
\usepackage{svg}

\usepackage{pifont}

\newcolumntype{P}[1]{>{\centering\arraybackslash}p{#1}}

\usepackage{array}
\newcolumntype{H}{@{}>{\setbox0=\hbox\bgroup}c<{\egroup}}

\usepackage{tikz}

\newcommand{\textct}[1]{\texttt{\textls[-30]{#1}}}

\usetikzlibrary{fit, positioning}

\begin{document}


\title{Poster: From Fort to Foe: The Threat of RCE in RPKI}

\author{Oliver Jacobsen}
\affiliation{ATHENE
    \city{Darmstadt}
    \country{Germany}
}
\affiliation{{Goethe-Universität Frankfurt}
    \city{Frankfurt}\country{Germany}
}

\author{Haya Schulmann}
\affiliation{ATHENE
    \city{Darmstadt}
    \country{Germany}
}
\affiliation{{Goethe-Universität Frankfurt}
    \city{Frankfurt}\country{Germany}
}

\author{Niklas Vogel}
\affiliation{ATHENE
    \city{Darmstadt}
    \country{Germany}
}
\affiliation{{Goethe-Universität Frankfurt}
    \city{Frankfurt}\country{Germany}
}

\author{Michael Waidner}
\affiliation{{ATHENE}
    \city{Darmstadt}
    \country{Germany}
}
\affiliation{{TU Darmstadt}
    \city{Darmstadt}
    \country{Germany}
}



\begin{abstract}
In this work, we present a novel severe buffer-overflow vulnerability in the RPKI validator Fort, that allows an attacker to achieve Remote Code Execution (RCE) on the machine running the software. 
We discuss the unique impact of this RCE on networks that use RPKI, illustrating that RCE vulnerabilities are especially severe in the context of RPKI. The design of RPKI makes RCE easy to exploit on a large scale, allows compromise of RPKI validation integrity, and enables a powerful vector for additional attacks on other critical components of the network, like the border routers.

We analyze the vulnerability exposing to this RCE and identify indications that the discovered vulnerability could constitute an intentional backdoor to compromise systems running the software over a benign coding mistake.
We disclosed the vulnerability, which has been assigned a CVE rated 9.8 critical (CVE-2024-45237).

\end{abstract}

\maketitle

\section{Introduction}

\tikzstyle{pp} = [rectangle, minimum width=.8cm, minimum height=.55cm,text centered, draw=black, fill=blue!30]
\tikzstyle{rp} = [rectangle, minimum width=.8cm, minimum height=.55cm,text centered, draw=black, fill=red!30]
\tikzstyle{is} = [rectangle, minimum width=.8cm, minimum height=.55cm, draw=black, fill=black!30, align=center]

\tikzstyle{arrow} = [thick,->,>=stealth]
\tikzstyle{darrow} = [thick,<->,>=stealth]
\tikzstyle{dashedarrow} = [thick,->,>=stealth, dashed]

\begin{figure}
\scriptsize
    \centering
    \begin{tikzpicture}
        \node (pp1) [pp, below=0, xshift=-3cm, yshift=-.2cm] {\textbf{PP1}};
        \node (pp2) [pp, below of=pp1, xshift=0cm] {\textbf{PP1}};
        \node (attacker) [below of=pp2, xshift=0cm] {
            \begin{minipage}{.75cm}
                \centering
                \includegraphics[width=0.75cm,trim={0.18cm 0.55cm 0.18cm 0.32cm},clip]{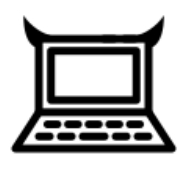}\\
                \textbf{Attacker}
            \end{minipage}
        };

        \node (rp) [rp, below=0, xshift=0cm, yshift=-1.2cm] {\textbf{RP}};
        \node (is) [is, below of=rp] {\shortstack{\textbf{Internal} \\ \textbf{Service}}};
    
        \node (router1) [below=0, xshift=3cm] {
            \begin{minipage}{.75cm}
                \centering
                \includegraphics[width=0.75cm,trim={0.18cm 0.56cm 0.18cm 0.9cm},clip]{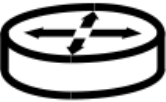}\\
                \textbf{Router}
            \end{minipage}
        };
        \node (router2) [below of=router1, xshift=0cm] {
            \begin{minipage}{.75cm}
                \centering
                \includegraphics[width=0.75cm,trim={0.18cm 0.56cm 0.18cm 0.9cm},clip]{res/pdf/router3.pdf}\\
                \textbf{Router}
            \end{minipage}
        };
        \node (router3) [below of=router2, xshift=0cm] {
            \begin{minipage}{.75cm}
                \centering
                \includegraphics[width=0.75cm,trim={0.18cm 0.56cm 0.18cm 0.9cm},clip]{res/pdf/router3.pdf}\\
                \textbf{Router}
            \end{minipage}
        };

        \draw[darrow] (rp) to [edge node={node [pos=0.55,below,sloped] {RPKI data}}] (pp1);
        \draw[darrow] (rp) to [edge node={node [pos=0.55,below,sloped] {RPKI data}}] (pp2);
        \draw[darrow] (rp) to [edge node={node [pos=0.55,below,sloped] {Payload}}] (attacker);

        \draw[darrow] (rp) to (router1);
        \draw[darrow] (rp) to (router2);
        \draw[darrow] (rp) to (router3);

        \draw[dashedarrow] (rp) to (is);
        
        \node[draw, dashed, inner xsep=0.15cm, inner ysep=0.02cm, fit=(rp) (is) (router1) (router2) (router3), xshift=-.05cm, label={[anchor=north west, xshift=0.1cm, yshift=-0.1cm]north west:Local Network}] (local) {};
    \end{tikzpicture}
    \vspace{-2.2em}
    \caption{Attack Setup.}
    \label{fig:setup}
    \vspace{-18pt}
\end{figure}

The Border Gateway Protocol (BGP) is the inter-domain routing protocol of the Internet.
The Resource Public Key Infrastructure (RPKI) was standardized to add security to BGP through cryptographic attestations, stored in distributed RPKI Publication Points (PPs).
Prefix owners upload Route Origin Authorizations (ROAs) to an RPKI PP.
Border routers use these ROAs for guiding their routing decisions in BGP.
The routers do not access PPs directly, instead, they install a middleware called Relying Party (RP) client that implements necessary RPKI functionality and provides the content to routers. RPs fetch objects from all the public RPKI PPs, parse and validate them, and compile a list of validated prefixed for the routers to use in BGP for best path selection.
This allows the routers to filter incoming BGP announcements for valid ownership of prefixes; if the announced Autonomous System (AS) number inside the BGP message differs from the authorized AS-prefix pair in the RPKI, the announcements are deemed invalid (i.e., may constitute a prefix hijack) and are hence rejected by the routers.

RPs are the central point of trust for BGP routers, since they are responsible for all parsing and validation functionality. Routers do not perform any additional cryptographic checks. This makes RPs an attractive target for attacks, as compromising their security potentially affects all the routers using them. Further, attacks against the RP may downgrade the RPKI protection, as the routers cannot use RPKI information for making routing decisions in BGP.

In this work, we identify and analyze a critical vulnerability in the RP software implementation Fort. We illustrate the technical details of the vulnerability, and discuss its impact in the context of RPKI and the deployment practices of RP clients.

In addition to the discovery and the technical analysis of this new RCE vulnerability, our work also provides novel insights into the devastating security implications of breaches of RP implementations. We develop an RFC attack vector exploiting the vulnerability that can be executed with low effort, and impacts all globally running RPs of Fort. We demonstrate that the attack provides a pathway to both compromise RPKI integrity, and potentially even attack border routers and other hosts in the local network.
 
We also discuss potential reasons for the existence of the buffer-overflow vulnerability. We derive factors from our analysis of this vulnerability that may indicate intent and not a ``bug.'' 
Amidst the tightened geopolitical and economical situations, the concerns for intentionally planted vulnerabilities, aka. backdoors, in software are growing and there are increasing indications for such backdoors, e.g., XZ Utils (CVE-2024-3094). We hope that our discussions make the community aware that RPs are also at risk of planted backdoors.

\section{RCE Vulnerability in Fort}
In our study into the security of RPKI RP implementations, we discovered a buffer-overflow vulnerability in the validation pipeline of Fort Validator.\newline
\indent \textbf{Vulnerable code.}
The vulnerability stems from a bug in the processing of the key-usage extension of an X.509 certificate, contained in most RPKI objects.
The 9 bit key-usage field indicates how the subject key of the certificate should be used, differentiating between keys to sign other Certificate Authority (CA) certificates, and keys to sign payload certificates. The key-usage extension of a validly formatted object has a value of \textit{0x0106} for CA certificate keys and \textit{0x0780} for payload certificates.

The vulnerability in the processing of the key-usage extension is in Listing \ref{lst:c-code}.
The function takes the bit string \texttt{ku} from the certificate and performs a memcpy operation, copying the extension value into the 2 byte \texttt{data} array located on the stack.
However, the number of copied bytes corresponds to the length of \texttt{ku}, which is controllable by the adversary and not checked. 
Therefore, the memcpy operation can overflow into the stack and write any sequential, attacker-controlled bytes into adjacent stack sections.

\textbf{Compilation.} The exploitability of the vulnerability depends on the compiler setup. 
Besides optimization passes, modern compiler toolchains, like LLVM and GCC, implement additional memory protections to prevent exploitation of memory bugs, like buffer overflows.
The (de)activation of these protection mechanisms is configured either directly by compiler flags in the build configuration or by a standard specification in the toolchain.
With the GCC toolchain, there are 3 options that impact how the code vulnerability translates into the compiled binary and therefore exploitability:

(1) Fortifications: When enabled, the memcpy calls are replaced with the memcpy\_chk built-in, which has runtime bounds checking. If \textct{ku->length} exceeds \textct{sizeof(data)}, the program exits. 
Fortifications can be disabled by undefining the macro \textct{\_FORTIFY\_SOURCE}.

(2) Built-ins: These function variants enable inline optimizations.
When memcpy is replaced with its built-in version, the compiler inlines and optimizes out the memcopy during compilation, as \texttt{data} is never read, therefore removing the vulnerable code from the final binary.
This can be disabled using \textct{-fno-builtin-memcpy}.

(3) Stack protector: The stack protector detects terminates the program if the stack is corrupted to protect against further exploitation. However, stack canaries are only applied to functions with local arrays under \textct{-fstack-protector-strong}, so supplying the \textct{-fstack-protector} flag will not include canaries in this function.

None of these protections are explicitly set in the build configuration.
Hence, an attacker only requires control of the default specification of the compiler toolchain to trigger the vulnerability to be compiled into the binary.
This can easily be achieved by injecting a single file into one of the compiler's search directories for spec files, or by manipulating the build environment and distributing the binary.
Instances compiled in older configurations, which do not have these protections enabled by default, are especially vulnerable.

\textbf{Exploitation.}
To exploit the vulnerability, the attacker creates a live RPKI repository, and inserts a malicious object.
The object triggers the vulnerability through a crafted key-usage extension.

Analysing the parsing and validation code of Fort, we find multiple requirements to the key-usage extension value to allow for exploitation.
First, the extension id, type, 
and length must be set to the correct value to prevent early validation failures. 
The internal field structure needs to resemble the expected values in the first three bytes of the value field.
The third byte must not contain any key-usage bits that are disallowed by the RPKI standard, effectively limiting the third byte to a value smaller than 0x8.

We observe experimentally that, when all requirements are met, the attacker can insert arbitrary additional bytes into the field value, allowing for a wide-range of inputs to place on the stack. 


\textbf{Escalation.}
Through placing arbitrary byte values on the stack, the attack can be escalated to Remote Code Execution (RCE).
We demonstrate a Proof-of-Concept (PoC) attack in our local setup without enabling compiler protections through a 13 byte payload that overwrites the function return pointer to a libc memory location, triggering Fort to open a local shell and allowing to execute arbitrary code. 

The attack can be generalized to other setups with enabled protections, depending on the target architecture and binary layout. For example, if the dynamic execution protection ASLR is active on the target machine, the attacker will have to spend additional effort to circumvent the protection.
This is generally possible, as the vulnerability allows for partially overwriting, e.g., writing over the return address pointer to redirect intra-binary control flow using brute-force. Further ways to circumvent ASLR can be found in \cite{evtyushkin2016jump}.

If full control over the execution is gained, the attacker may access the machine, e.g., through a remote-shell payload inserted into the Fort control-flow.
Gaining access to the RP system enables a range of attacks, including attacks on RPKI security, host security, and even attacks on the local network of the system using the RP, which we elaborate in Section~\ref{sec:evimp}.

\begin{figure}[t!]
    \centering
    \begin{lstlisting}[language=C, caption={Vulnerable function.}, label={lst:c-code}, 
    basicstyle=\ttfamily\footnotesize,   % Smaller font size
    keywordstyle=\color{blue},           % Color for keywords
    commentstyle=\color{gray},           % Color for comments
    numberstyle=\tiny\color{gray},       % Style for line numbers
    numbers=left,                        % Line numbers on the left
    stepnumber=1,                        % Line number step
    numbersep=5pt,                       % Space between line numbers and code
    %backgroundcolor=\color{lightgray!10},% Light gray background color
    frame=single,                        % Single frame around the code
    frameround=tttt,                     % Rounded corners
    tabsize=4,                           % Tab size
    showspaces=false,                    % Do not show spaces
    showstringspaces=false,              % Do not show string spaces
    breaklines=true,                     % Break long lines
    breakatwhitespace=true,              % Break at whitespace
    escapeinside={(*@}{@*)},             % Escape inside (*@ @*)
    language=C                           % Language
]
static int
handle_ku(ASN1_BIT_STRING *ku, unsigned char byte1)
{
    unsigned char data[2];
    memset(data, 0, sizeof(data));
    memcpy(data, ku->data, ku->length);   
    if (ku->data[0] != byte1) {
        return pr_val_err("Illegal flag [...]");}  
    return 0;}
\end{lstlisting}
\vspace{-15pt}
\end{figure}

\textbf{Real-world exploitation.} 
Exploitation of the buffer overflow on a real-world setting can be achieved by placing a malicious RPKI object inside a live RPKI PP, and getting the victim RP to download and process the object.
This setup can be achieved by a small-scale attacker. Setting up a RPKI repository requires ownership of Internet resources and membership in one of the five Regional Internet Registries, both of which only require minor financial investment and administrative overhead.
After setting up their repository, all global RPs will regularly download and validate the objects from the attacker, giving the attacker access to all globally running RP instances of Fort, which makes wide-spread exploitation possible.
Further, since the attack can be constructed to allow the RP to continue processing after executing the payload, the attack can be conducted stealthy.
The RP will only log a benign error about a malformed key-usage extension, making detection difficult. Exploitation of the vulnerability can thus be executed with moderate setup effort, and can attack any globally running instance of Fort. 


\section{Evaluation and Impact}\label{sec:evimp}
The buffer overflow vulnerability can have devastating consequences for systems running the vulnerable RP.

\textbf{Evaluation.} We find that the vulnerable code was added to the repository in January 2019. Therefore, all Fort version from Beta-version 2 to version 1.6.2 are vulnerable to buffer overflow. We confirm vulnerability of the most recent version 1.6.2 by setting up a local isolated repository with the RPKI repository tool CURE~\cite{mirdita2023cure} and inserting a manipulated key-usage bitstring. We observe that in a default compiler configuration, Fort crashes with a buffer overflow warning, triggered by the stack protector. 

To quantify the amount of systems that could be affected by this vulnerability, we setup our own well-configured live RPKI PP and log the amount of Fort clients with the vulnerable version connecting to us. We find over 100 systems running a vulnerable version of Fort, including major systems, like Amazon or Internet Systems Consortium, and a subset of RIRs, all of which may be attacked through the presented exploit, depending on their utilized compiler toolchain. 
Any systems that compiled Fort with a toolchain not enforcing necessary protections to prevent exploitation of the bug are vulnerable.
To run our experiments ethically, we do not test the exploit on any system contacting our PP. Observing the large amount of systems that can be attacked with RCE, we additionally provide evaluation of the potential impact of the attack on hosts running the vulnerable client. 

\textbf{RPKI impact.} 
Compromising the RP through a RCE exploit allows arbitrary manipulations of RPKI data. Since the data provided to routers does not require any signatures, the attacker {\em does not need to forge any signatures to get the manipulated data accepted}. This allows creation of malicious route origins to enable hijacking of arbitrary prefixes. Since the data from the RP is trusted, the router will validate the hijacking prefix announcement through the malicious RPKI data. Vice versa, the attacker can get arbitrary prefix announcements invalidated by deleting or changing the appropriate RPKI data, 
leading to announcements classified as RPKI invalid and hence dropped. 
With continuous deployment of new technologies within RPKI, the specific impact will further increase.

\textbf{Host impact.} Exploiting the RCE gives the attacker access to the host, where other services may run or sensitive files might be stored. Potential host-level attacks include stealing sensitive data, like user-passwords, stored credentials, or SSH private keys, installation of malware, or DoS.

\textbf{System impact.}
RPs provide data to BGP routers and must therefore run in a part of the network that has access to the routers. This makes the machine running the RP a valuable entry-point into the local network.
Discovery of reachable BGP routers within the network is straightforward, as all routers utilizing RPKI must regularly poll the RP for updates, allowing an attacker easy enumeration.
The attacker can use the RP machine to launch additional attacks on BGP routers, the management interfaces of which are otherwise isolated from external traffic.
The attacker can stuff credentials, exploit known vulnerabilities, or run a DoS attack on the routers. Further, since RPs are software components that generally fall within the category of network management infrastructure, it is sensible that the local network of the RP might contain additional interesting targets for the attacker, like management servers.

Exploiting the RCE on an RP client thus not only compromises RPKI security, but also enables a vector for further attacks on the host, and other hosts and routers within the local network.

\balance

\textbf{Source of vulnerability.} Observing the devastating impact of the vulnerability raises questions on how the bug was introduced to the code. We find potential indications for both intentional planting, and a benign coding error. First, the code section does not implement any functionality, putting into question why the code was added in the first place. Further, the convenient accessibility of the vulnerability from any remote PP makes exploitation easy and stealthy. Also, the vulnerability can be activated in new software distributions by simply changing compilation flags, which will likely not be noticed by users. All these observations might indicate malicious intent. However, we do not find any operational indications for malicious planting of the vulnerability. The code was committed by the long-term main developer of Fort, and similar code sections exist in other parts of the RP that implement actual functionality, making a copy-paste error likely. While there is no clear answer on the intention behind adding the code, and it might very well constitute a benign coding error, the ease with which such a backdoor could be inserted should raise the awareness of users when using any software communicating with remote servers. 
\section{Conclusions}\label{sc:conc}
In this work, we presented a new severe vulnerability in RPKI software, enabling RCE on a victim machine running the RP client Fort.
Following our analysis of the impact of the vulnerabilities in RPKI software, and considering the abundance of vulnerabilities in RPKI~\cite{mirdita2023cure} and the relatively low effort for large-scale exploitation, we argue that the security of RPKI will get increasingly important in the near future, and the community might see a stark increase in malicious attacks on the architecture. Systems running RPKI need to be aware of the potential impact of attacks and need to strive for improved security of implementations, and for operational practices that ensure security of the system if the RP is compromised. This includes not storing any sensitive data, including router credentials, on RP servers, implementing separation of the local network, and setting up monitoring solutions that can detect compromises to the security of the RP. We hope that our work, including the responsible disclosure of the vulnerability, aids in improving the security of RPKI implementations, and sensitizing the community for the increasing thread to the technology in the near future.

\section*{Acknowledgements}
This work has been co-funded by the German Federal Ministry of Education and Research and the Hessian State Ministry for Higher Education, Research and Arts within the National Research Center for Applied Cybersecurity ATHENE and by the DFG, German Research Foundation SFB~1119.

\bibliographystyle{ACM-Reference-Format}
\bibliography{ref,bib}

\end{document}